\newcommand{\sam}[1]{{\small\color{orange}{Samal:#1}}}

\documentclass[acmsmall,screen]{acmart}

\usepackage{amsmath}
\usepackage{amssymb}
\setcopyright{none}
\renewcommand\footnotetextcopyrightpermission[1]{} 
\usepackage{tikz}
\usepackage{subcaption}
\usepackage{multirow}
\usepackage{booktabs}
\usepackage{amsmath, amssymb}
\usetikzlibrary{arrows.meta, positioning}
\usepackage{tcolorbox}
\usepackage{subcaption}
\usepackage{tabularx}
\usepackage{array}
\usepackage{hyperref}

\AtBeginDocument{%
  }

\usepackage{xspace}
\newcommand{\tool}{\textsc{VulReaD}\xspace}
\newcommand{\commentout}[1]{}
\begin{document}
\pagestyle{empty}
\thispagestyle{empty}
\title{\tool: Knowledge-Graph–guided Software Vulnerability Reasoning and Detection}


\author{Samal Mukhtar}
\affiliation{%
  \institution{The University of Manchester}
  \city{Manchester}
  \country{United Kingdom}}

\author{Yinghua Yao}
\affiliation{%
  \institution{Agency for Science, Technology and Research}
  \city{Singapore}
  \country{Singapore}
}

\author{Zhu Sun}
\affiliation{%
 \institution{Singapore University of Technology and Design}
 \city{Singapore}
 \country{Singapore}}

\author{Mustafa Mustafa}
\affiliation{%
  \institution{The University of Manchester}
  \city{Manchester}
  \country{United Kingdom}}

\author{Yew Soon Ong}
\affiliation{%
  \institution{Agency for Science, Technology and Research}
  \city{Singapore}
  \country{Singapore}}

\author{Youcheng Sun}
\affiliation{%
  \institution{Mohamed Bin Zayed University of Artificial Intelligence}
  \country{United Arab Emirates}}


\begin{abstract}

Software vulnerability detection (SVD) remains a critical challenge in the security of modern systems.  Large language models (LLMs) have recently been explored for SVD because they can provide natural-language explanations alongside predictions. However, most previous work still focuses on binary evaluation, and the resulting explanations often contain hallucinations or lack semantic consistency with Common Weakness Enumeration (CWE) categories, limiting their reliability in practice—particularly when ground-truth validation and expert supervision are expensive.
To address this gap, we propose \textbf{\tool}, a {k}nowledge-{g}raph–guided approach for \underline{vul}nerability \underline{rea}soning and \underline{d}etection that moves beyond binary classification toward CWE-level reasoning. \tool leverages a security knowledge graph (KG) as a shared semantic backbone and employs a strong teacher LLM to generate CWE-consistent contrastive reasoning supervision, enabling scalable, semantically guided student model training without manual preference annotation. Student models are then adapted using parameter-efficient fine-tuning combined with Odds Ratio Preference Optimization (ORPO), encouraging faithful, taxonomy-aligned reasoning while suppressing misleading or unsupported explanations.
Through a comprehensive study across three real-world datasets, \tool achieves consistent gains in both CWE-level reasoning and binary vulnerability detection. On average, it improves binary F1-scores by around 8\% to 10\% and increases multi-class classification performance by approximately 30\% Macro-F1 and 18\% Micro-F1 compared to state-of-the-art baselines through better CWE-level reasoning. These improvements confirm robustness across benchmarks and highlight two key insights: (i) LLMs generally outperform deep learning baselines in binary detection, and (ii) KG-guided reasoning substantially enhances accuracy and CWE coverage in multi-class detection, yielding more reliable and interpretable vulnerability analysis.

\end{abstract}



\maketitle
\thispagestyle{empty}
\section{Introduction}


Software vulnerabilities remain a persistent and critical threat to modern software systems, enabling attackers to compromise system integrity, leak sensitive information, and disrupt essential services. Despite decades of research on automated software vulnerability detection (SVD), vulnerabilities continue to appear in real-world software at scale, motivating continued interest in improving detection techniques.

Researchers have developed various SVD methods that mainly rely on pattern-based static analysis or deep learning (DL) models trained on historical vulnerability data \cite{chakraborty2021deep, li2021sysevr, li2018vuldeepecker}. Although these methods have achieved measurable success, they often struggle to generalize to diverse vulnerability patterns and evolving codebases because they learn superficial patterns over meaningful vulnerability indicators \cite{ding2024vulnerability}. 
More recently, large language models (LLMs) have emerged as a promising alternative for SVD due to their strong code comprehension along with generative capabilities \cite{zibaeirad2025reasoning, hou2024large}. By leveraging large-scale pretraining and fine-tuning strategies, LLM-based approaches have demonstrated encouraging results in detecting vulnerabilities \mbox{at the function level~\cite{shestov2025finetuning}.}

\paragraph{Why Existing Detection is Not Enough} However, existing evaluations of SVD methods are largely dominated by binary classification settings, where models are tasked only with distinguishing vulnerable from non-vulnerable code. Although binary detection provides a convenient benchmark, it fails to capture whether a model truly understands the nature of a vulnerability \cite{labruna2025retrieve}. In practice, effective vulnerability mitigation requires identifying the underlying weakness, typically expressed as a Common Weakness Enumeration (CWE) category. As a result, binary accuracy alone may substantially overestimate the real-world utility of SVD methods. 

Although a small number of previous studies explore multi-class vulnerability classification, they typically restrict the task to a limited subset of frequent CWE types \cite{wen2024livable, lekssays2025llmxcpg,du2024vul,cao2024snopy,yu2025preliminary}. Such settings do not reflect the long-tailed and diverse nature of real-world vulnerabilities and obscure the scalability challenges faced by both DL and LLM-based approaches \cite{deng2024improving, jiang2024investigating}. Consequently, it remains unclear how well existing models perform when required to reason \mbox{in a wider CWE space~\cite{weyssow2025r2vul}.}

\paragraph{From Detection to Diagnosis} Moreover, recent LLM-based methods increasingly emphasize natural language explanations or reasoning to justify detection decisions \cite{khare2025understanding}. While these explanations often appear coherent and convincing, their semantic alignment with the correct CWE categories has not been rigorously examined \cite{weyssow2025r2vul}. This raises an important question: \textit{Does enhanced reasoning genuinely reflect a deeper understanding of vulnerabilities, or does it simply obscure underlying misclassification errors?} 

\paragraph{Approach Overview} In this work, we investigate whether LLM reasoning can improve multi-class vulnerability detection across diverse CWE categories. Our central idea is to evaluate not only classification accuracy but also the semantic alignment of model reasoning with ground-truth vulnerability types. Addressing this objective involves several challenges. The distribution of CWE categories is highly imbalanced, with a small number of frequent types and a long tail of rare vulnerabilities. Extending evaluation beyond these frequent categories introduces scalability concerns for both DL- and LLM-based approaches. In addition, while LLMs often generate fluent reasoning, it remains uncertain whether such reasoning consistently aligns with correct vulnerability types, raising questions about reliability in practical settings.


Inspired by the recent success of Knowledge Graphs (KG) in domains such as question answering~\cite{wu2025graph, wang2025knowledge, wang2024learning} and impact of external knowledge in model performance \cite{labruna2025retrieve, feng2024don,zhang2023retrieve,simoni2025morse}, we propose \textbf{\tool}, a {k}nowledge-{g}raph–guided approach for \underline{vul}nerability \underline{rea}soning and \underline{d}etection. Our work is motivated by the unique challenges of software vulnerability analysis, where objective correctness is critical and high-quality supervision is costly to obtain. Unlike many natural language tasks, vulnerability reasoning requires precise identification of underlying security causes and accurate CWE attribution, while large-scale, expert-annotated reasoning data remains scarce and expensive to construct \cite{croft2023data, wen2023less}. 

These constraints make direct human preference annotation or test-based feedback impractical, particularly across the long-tailed CWE space. As a result, strong teacher models emerge as a practical and scalable mechanism for providing structured, semantically grounded supervision. Recent advances in reasoning-aware vulnerability detection further support this design choice: R2Vul~\cite{weyssow2025r2vul} demonstrates that training LLMs with Odds Ratio Preference Optimization (ORPO) over contrastive reasoning pairs yields state-of-the-art performance for binary vulnerability detection, providing strong empirical evidence that explicit reasoning supervision improves detection accuracy. However, this line of work remains focused on binary classification and does not explicitly address multi-CWE coverage or evaluate the semantic consistency of generated reasoning with vulnerability taxonomies. Consequently, models may achieve high detection accuracy while still producing incomplete or weakly grounded explanations at the CWE level.

Building on these insights, \tool is designed to move beyond binary vulnerability detection and explicitly target CWE-level reasoning, where correct classification depends on understanding vulnerability semantics rather than surface code patterns. At its core, \tool employs a security KG that organizes vulnerability knowledge by relating CWEs through common code entities and conceptual security abstraction classes \cite{li2025safegenbench}. This KG captures both hierarchical and semantic relationships among vulnerabilities and serves as a unifying structure for dataset construction, teacher–student training, and inference-time guidance.
First, we construct an enriched dataset by augmenting real-world benchmarks with KG-derived attributes. A strong teacher LLM generates contrastive reasoning pairs for each example, consisting of a CWE-aligned reasoning and a deliberately flawed counterpart. The CWE-aligned reasoning highlights faulty code constructs, discusses vulnerability mechanisms and their impact, and contextualizes them with vulnerability behavioral context. The flawed reasoning is produced through label flipping, yielding explanations that are inconsistent with the target CWE. These paired outputs differ not only in fluency but in their semantic consistency with the KG-defined CWE mappings. We then adapt student LLMs using parameter-efficient fine-tuning combined with (ORPO). By optimizing both supervised prediction accuracy and preference alignment, ORPO encourages the student to favor factually grounded, CWE-consistent reasoning while suppressing misleading or hallucinated explanations that frequently arise in unconstrained LLM outputs.

\paragraph{Key Findings and Contributions} We evaluate our approach through a comprehensive empirical study comparing DL models and LLM-based methods under both binary and multi-class SVD settings. Our best-performing \tool based on Qwen2.5-7B-Instruct achieves consistent gains over state-of-the-art methods in both CWE-level reasoning and binary SVD. On the widely used PrimeVul dataset \cite{ding2024vulnerability}, \tool improves F1-scores by 4\%, 11\%, and 10\% compared to leading binary classifiers.
To contextualize these results, we evaluate three DL models and three LLM families (LLaMA, Qwen, DeepSeek) under multiple adaptation techniques. DL models achieve only about 30\% average F1-scores, which is substantially lower than LLM baselines. In CWE-level reasoning, \tool outperforms the reasoning-based baseline by 30\% in macro-F1 and 18\% micro-F1 on average, and consistently surpasses LLM baselines across datasets.
These findings demonstrate two key insights: (i) LLMs generally outperform DL models in binary SVD tasks, and (ii) KG-guided reasoning substantially enhances accuracy and CWE coverage in multi-class detection, yielding more reliable and interpretable vulnerability analysis.

This paper makes the following contributions:
\begin{itemize}
    \item \textbf{CWE-level vulnerability reasoning and detection.} We propose \tool, an approach that moves beyond binary detection to CWE-level diagnosis over a long-tailed vulnerability space.
    \item \textbf{Knowledge-guided dataset distillation.} We augment existing benchmarks (PrimeVul~\cite{ding2024vulnerability} , DiverseVul~\cite{chen2023diversevul}, R2Vul~\cite{weyssow2025r2vul}) with structured attributes and construct paired CWE-consistent and CWE-inconsistent rationales, supporting more consistent and interpretable vulnerability reasoning. 
    \item \textbf{Preference optimization for grounded explanations.} We combine parameter-efficient fine-tuning with ORPO to improve the factual grounding and CWE consistency of LLM-generated vulnerability explanations.
    \item \textbf{Empirical study of reasoning and detection.} We systematically evaluate various DL models and LLMs under both binary and multi-class vulnerability detection settings, showing that strong binary performance often does not translate into accurate CWE identification or semantically correct explanations.
\end{itemize}

\commentout{
\begin{itemize}
    \item \textbf{KG-guided vulnerability detection and reasoning}. We propose a knowledge-graph–guided approach for software vulnerability detection that integrates structured security knowledge (CWEs, security abstractions, and entity-level relations) into data construction, training, and inference, enabling CWE-level reasoning over a long-tailed vulnerability space.
    \item \textbf{KG-informed dataset distillation}. We distill existing datasets (PrimeVul, DiverseVul, R2Vul) by constructing paired true and false reasonings guided by KG information. The resulting dataset encodes structured knowledge with function-level entities, security abstract classes, and CWE mappings, providing a foundation for consistent and interpretable vulnerability reasoning. \sam{Updated}
    \item \textbf{KG-informed preference optimization for LLM reasoning}. Along with dataset distillation we introduce a training strategy that combines parameter-efficient fine-tuning with ORPO to improve the factual grounding and CWE consistency of LLM-generated vulnerability explanations.
    \item \textbf{Empirical study on reasoning and detection}. We conduct a systematic comparison of deep learning models and large language models under both binary and multi-class vulnerability detection settings, showing that binary accuracy substantially overestimates CWE-level reasoning capability.
\end{itemize}
}
\section{Background and Motivation}

\subsection{Software Vulnerability Detection}
SVD aims to identify defects in source code that may be exploited to compromise system security. With the increasing availability of labeled vulnerability datasets DL techniques have been widely adopted for vulnerability detection \cite{chakraborty2021deep}. Most prior learning-based approaches operate on function-level or statement-level code representations and formulate vulnerability detection as a binary classification task \cite{russell2018automated, rahman2024towards}. While these models have demonstrated competitive detection performance, they typically provide limited interpretability and do not explain the underlying causes of detected vulnerabilities \cite{ni2023distinguishing, zhou2024comparison}.
More recently, LLMs pretrained on large-scale code corpora have been explored for vulnerability detection. Because of their strong language modeling capabilities, LLMs can analyze code semantics and generate natural-language explanations \cite{wang2024llmdfa, sun2024llm4vuln, sultana2024code}. This capability suggests a potential shift from purely predictive models toward systems that can assist developers by explaining vulnerability causes. However, the reliability and correctness of such explanations remain an open challenge~\cite{ullah2024llms, haroon2025accurately}.

\subsection{Vulnerability Taxonomies and Security Knowledge}
To support consistent vulnerability reporting and analysis, the security community has developed standardized vulnerability taxonomies and structured knowledge resources \cite{barnum2012standardizing, fenz2009formalizing, syed2016uco}. The Common Vulnerability Enumeration (CVE\footnote{https://cve.mitre.org/}) system assigns unique identifiers to publicly disclosed vulnerabilities in specific software products. These identifiers are widely used to aggregate vulnerability information across databases and tools.

Complementing CVE, the CWE\footnote{https://cwe.mitre.org/} provides a hierarchical taxonomy of abstract software weakness types that characterize recurring vulnerability causes, such as improper input validation or insecure file handling. CWE focuses on the root causes of vulnerabilities at the design or implementation level, rather than individual vulnerability instances.

CWE and CVE are often integrated into larger security databases, such as the National Vulnerability Database (NVD\footnote{https://nvd.nist.gov}), which associates CVE entries with severity metrics and descriptive metadata~\cite{host2023constructing}. In this work, we focus on CWE as the primary semantic abstraction, as CWE categories provide a meaningful basis for reasoning about vulnerability causes and for supporting remediation and secure development practices.

\subsection{Knowledge Graphs for Security Representation}
A KG is a structured representation of entities and their relationships, typically modeled as a graph in which nodes denote entities and edges encode semantic relations. KG captures hierarchical and relational information and can incorporate rich attributes associated with entities and relationships.

In the security domain, KGs provide a natural mechanism for organizing vulnerability-related knowledge, including relationships between CWE categories, vulnerability descriptions, affected components, and mitigation concepts \cite{shi2024uncovering, mouiche2025entity, wu2025ng_mderank, host2023constructing}. Such representations enable structured querying and reasoning over security knowledge and offer a principled way to anchor automated analyzes in standardized taxonomies \cite{liu2025seckg2vec, dong2024intelligent}.

To manage the complexity of KG data, we employ the node-based database Neo4j\footnote{https://neo4j.com/}. As one of the most widely used graph databases, Neo4j provides a scalable, reliable, and secure platform for storing, querying, and analyzing large volumes of interconnected information. Its robust graph storage and processing capabilities, together with the expressive Cypher query language, enable efficient development, deployment, and maintenance of graph-based applications across projects of varying scale.

\subsection{Motivation and Challenges}
\label{motivsec}

Recent LLM-based vulnerability detection systems show promising performance \cite{steenhoek2024comprehensive}, but our preliminary analysis reveals consistent reasoning failures. Even when they correctly flag vulnerable code, their explanations are often misleading and poorly aligned with CWE categories, especially for rare vulnerability types, where models default to common or semantically adjacent CWEs.

\begin{figure}[!htp]
  \centering
  \includegraphics[width=\linewidth]{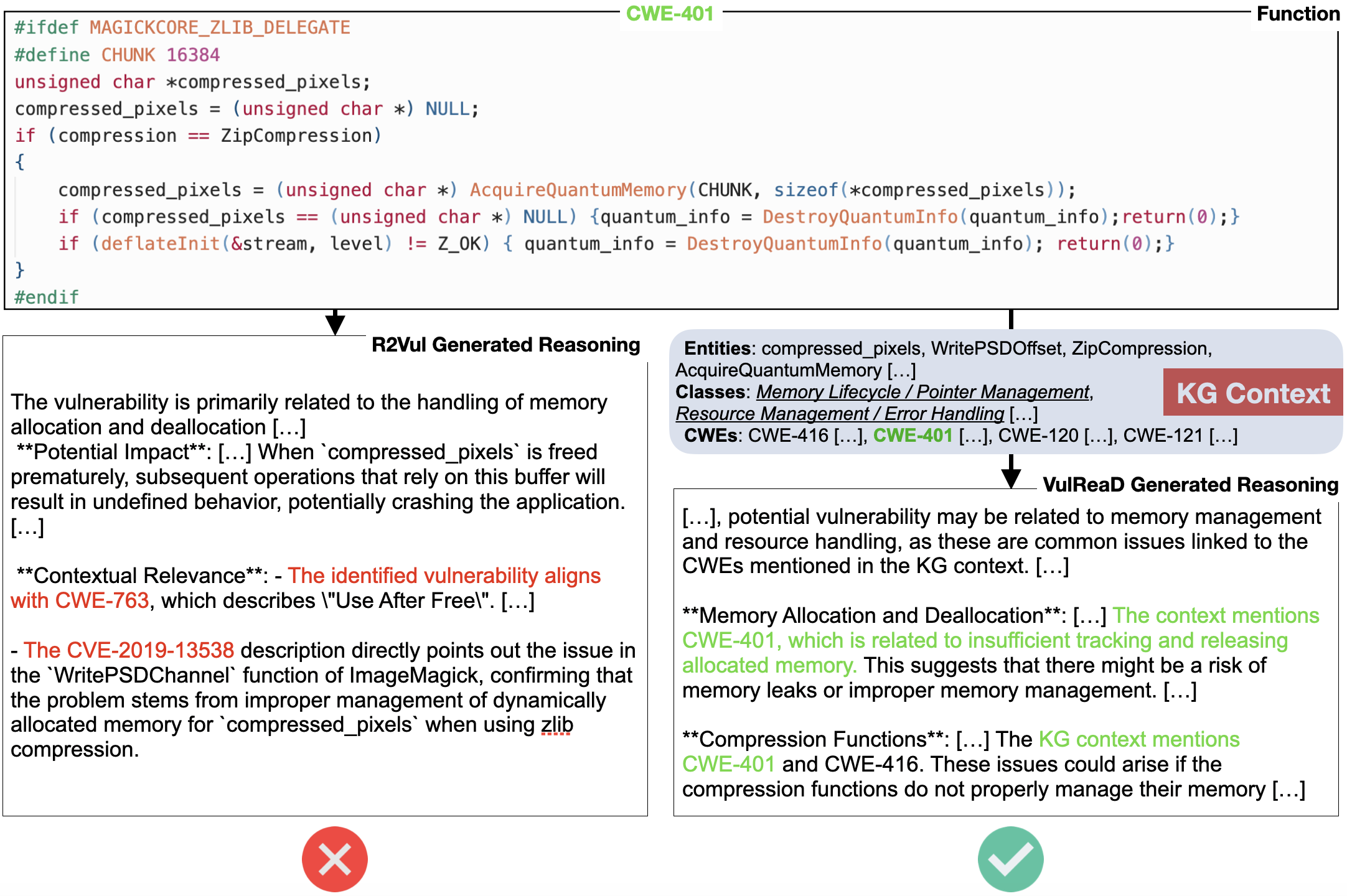}
  \caption{Failure case of baseline R2Vul on CWE-level reasoning. Given a function labeled CWE-401 (Memory Leak) (top), R2Vul generates a fluent rationale that incorrectly attributes the issue to CWE-763 (Use-After-Free) and introduces an unrelated CVE reference (bottom), illustrating semantic misalignment between generated explanations and the target CWE.}
  \Description{...}
  \label{motiv}
\end{figure}

\begin{figure}[h]
  \centering
  \includegraphics[width=\linewidth]{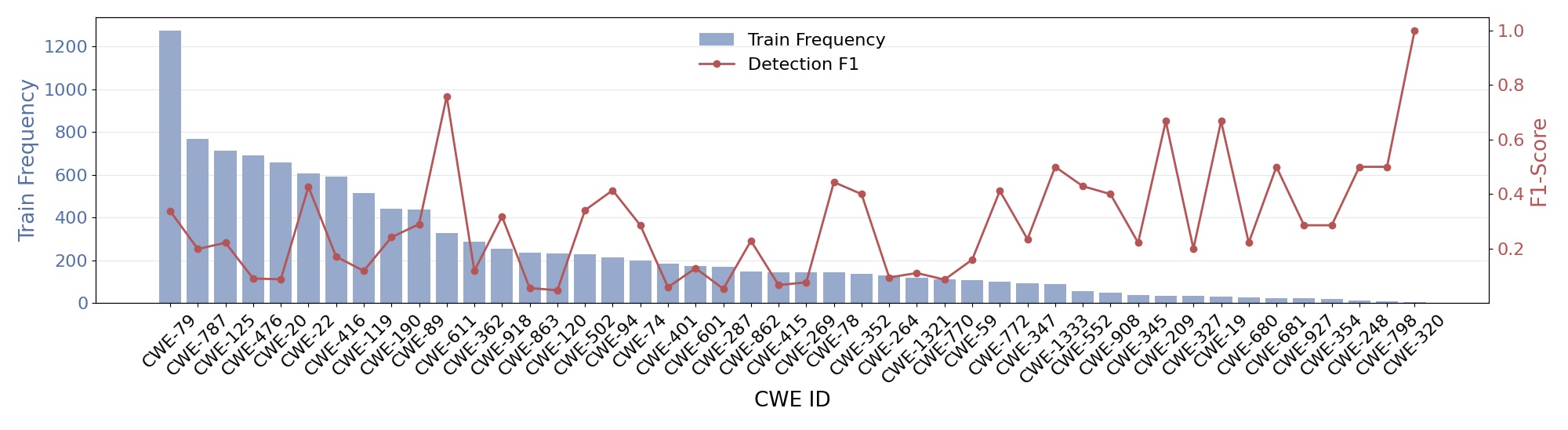}
  \caption{Training CWE Data Frequency vs. Reasoning-Based CWE Identification Accuracy.}
  \Description{...}
  \label{motiv2}
\end{figure}

Figure \ref{motiv} illustrates the reasoning limitation of the baseline R2Vul approach using an example from the R2Vul dataset. Without external knowledge guidance, the LLM generates a vague explanation and an incorrect CWE attribution to CWE-763 instead of ground-truth CWE-401, despite making a correct binary detection. This suggests that detection accuracy alone is insufficient to ensure reliable and trustworthy vulnerability analysis.
Notably, this limitation persists even when the model is trained using ORPO, which optimizes preference alignment between valid and flawed reasoning outputs. While ORPO improves surface-level coherence, it does not guarantee that the model correctly identifies the underlying vulnerability cause within the function. As a result, the model may assign an incorrect or weakly related CWE, and in some cases exhibit confusion between CWE and CVE identifiers.
A direct comparison with a \tool generated reasoning further emphasizes this issue. When the same example is analyzed using our KG-guided approach, the model correctly identifies CWE-401 as the primary vulnerability and produces a coherent explanation grounded in the root cause. 
To illustrate the quantitative limitation of this gap, Fig. \ref{motiv2} summarizes reasoning performance, showing low CWE-level reasoning performance in both coverage and accuracy.

These observations highlight a critical gap in current SVD research: most approaches optimize for binary correctness while neglecting the reliability and semantic consistency of model-generated explanations. This gap motivates the need for systematic evaluation of LLM reasoning quality and mechanisms that guide vulnerability explanations to structured security knowledge. In particular, integrating CWE-centered knowledge representations such as KG offers a promising direction for improving the trustworthiness and interpretability of LLM-based vulnerability detection systems.

\section{Problem Formulation}
\subsection{Task Definition}
We study the problem of SVD with \emph{semantically correct} vulnerability reasoning. Let $f$ denote a function-level code snippet (a sequence of tokens or lines). Given $f$, many existing SVD systems predict a binary label $y \in {0,1}$, where $y=1$ indicates that $f$ contains at least one vulnerability.

Beyond binary detection, we consider \emph{vulnerability cause identification}. Let $C$ denote the set of CWE categories organized as a taxonomy. For a vulnerable function ($y=1$), the SVD system is expected to predict one or more CWE IDs $c \in C$ that characterize the underlying weakness.

\subsection{LLM-Based Vulnerability Reasoning}
Our work follows recent approaches that use LLMs to analyze $f$ and produce predictions together with a natural-language explanation. Formally, given a function $f$, an LLM produces: 1) a vulnerability prediction $y$, and 2) a textual explanation $r$ intended to justify the prediction.

Ideally, the explanation $r$ should accurately describe the root cause of the vulnerability and align with the predicted CWE-ID $c$. However, in practice, explanation correctness is not guaranteed. An LLM may identify the presence of a vulnerability while mischaracterizing its cause, or may assign a CWE label that is only weakly related to the actual weakness in the code.
This mismatch motivates the need to explicitly model and evaluate the relationship between predictions, generated explanations, and standardized vulnerability semantics.

\subsection{KG-based Vulnerability Reasoning}
To capture structured vulnerability knowledge, we assume access to a security KG, $G = \{V, E\}$, where:
\begin{itemize}
    \item nodes $V$ represent security entities such as CWE categories, security abstractions, and descriptive concepts, and
    \item edges $E$ encode semantic relations, including hierarchical (e.g., parent–child) and associative relationships.
\end{itemize}
Each CWE node is associated with textual descriptions and metadata that characterize the corresponding vulnerability type. The KG provides an explicit semantic space in which relationships between vulnerability concepts can be queried and reasoned about.
\subsection{Problem Statement}
Given a function-level code snippet $f$, an LLM $M$, and a security KG $G$ centered on CWE semantics, our objective is to design a vulnerability detection approach that:
\begin{enumerate}
    \item correctly predicts whether $f$ is vulnerable,
    \item identifies the most appropriate CWE category(ies) describing the vulnerability cause, and
    \item produces explanations that are semantically anchored in the KG and consistent with the identified CWE.
\end{enumerate}
Crucially, we focus not only on a prediction accuracy, but also on the faithfulness and interpretability of the model’s reasoning. A correct vulnerability prediction accompanied by an incorrect or misleading explanation is considered a failure in this setting.

\subsection{Challenges}
This problem presents several key challenges:
\begin{itemize}
    \item \textbf{Semantic ambiguity}: Source code may exhibit symptoms that correspond to multiple related CWE categories, making precise cause identification difficult.
    \item \textbf{Long-tailed label space}: CWE categories are unevenly distributed, with many fine-grained weakness types appearing infrequently in training data.
    \item \textbf{Unconstrained reasoning}: LLM-generated explanations are free-form and may rely on spurious cues or hallucinated security concepts.
    \item \textbf{Misalignment between detection and explanation}: High binary detection accuracy does not guarantee correct vulnerability attribution at the CWE level.
\end{itemize}
These challenges motivate approaches that explicitly integrate structured security knowledge into both training and inference, enabling LLMs to reason within a constrained and interpretable semantic space.

\section{\tool} 

\begin{figure}[h]
  \centering
  \includegraphics[width=\linewidth]{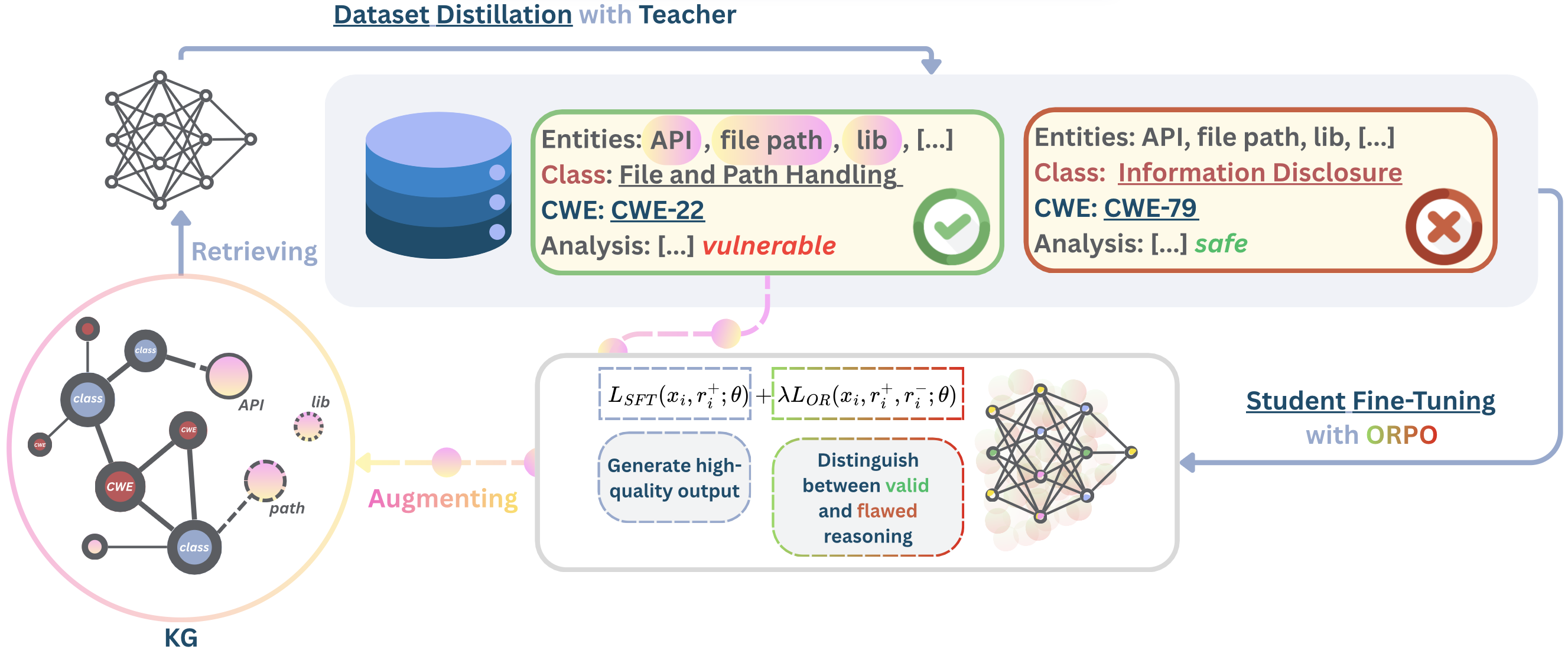}
  \caption{Overview of \tool. The security knowledge graph (KG) provides structured vulnerability semantics (entities, abstraction classes, and CWE relations) that are retrieved to enrich each training example. A teacher LLM distills the dataset by generating a \emph{contrastive} rationale pair: a CWE-consistent (valid) analysis and a deliberately CWE-inconsistent (flawed) analysis. A student model is then fine-tuned with ORPO, combining supervised fine-tuning on valid rationales ($L_{\text{SFT}}$) with an odds-ratio preference term ($L_{\text{OR}}$) to favor grounded, CWE-aligned reasoning over flawed alternatives.}
  \Description{...}
  \label{approach}
\end{figure}

To address these challenges, we propose \tool, a knowledge-graph--guided approach for vulnerability reasoning and detection that integrates structured security knowledge into both LLM training and inference for explainable SVD. Figure~\ref{approach} provides an overview of the \tool workflow.  Given a labeled vulnerability dataset, we first construct a structured preference dataset by enriching each function with valid and flawed reasoning generated by a teacher LLM based on the custom KG. Therefore, unlike prior work that focuses on free-form natural language explanations~\cite{weyssow2025r2vul}, we extract explicit entities from functions mapped to security abstract classes.
We then fine-tune a student model using ORPO to prefer structurally consistent reasoning over flawed alternatives. During inference, the KG serves as an external semantic guide, enabling CWE-consistent and security-aware predictions.

\subsection{Security KG Construction with CWE Abstraction}
To enable a more accurate and comprehensive evaluation of LLM-generated code with respect to security vulnerabilities, the benchmark dataset should exhibit sufficient diversity across both programming language paradigms and vulnerability categories, which ensure broad applicability and support a rigorous assessment of code robustness. 
Inspired by \cite{li2025safegenbench}, we first construct a taxonomy of common software vulnerabilities by introducing 13 conceptual vulnerability abstraction classes, each of whom reflects a specific class of security flaws (e.g., File and Path Handling, Input Validation, Access Control, Memory Management). These are designed to capture both underlying mechanisms of vulnerabilities and the conceptual patterns commonly used when reasoning about security flaws in code. Thus, it provides an interpretable abstraction layer over hundreds of fine-grained CWE labels, forming the structural backbone of the KG used throughout training and inference. All CWEs in the official CWE corpus are covered by at least one abstract class, ensuring comprehensive vulnerability coverage.

To associate CWEs with abstract classes, we employ a hybrid mapping strategy that combines symbolic and semantic signals. In \textit{Keyword-based Mapping}, each class is defined by a curated set of security-related keywords reflecting its vulnerability mechanism. The CWEs descriptions are first scanned for these keywords, and the CWEs containing explicit matches are assigned directly to the corresponding abstract class. For the CWEs without direct keyword matches, we employ an \textit{Embedding-based Mapping} where we compute sentence embeddings of CWE descriptions using a SentenceTransformer model. These embeddings are compared against pre-computed class description embeddings using cosine similarity, and the CWE is assigned to the class with the highest similarity score. Because many vulnerabilities exhibit overlapping characteristics, CWEs are allowed to map to multiple abstract classes. This design preserves the multidimensional nature of security weaknesses rather than forcing overly coarse categorization.

\subsection{KG-Grounded Data Construction and Reasoning Distillation}
\label{distill}
To construct training data that explicitly reflects security knowledge, we introduce a KG-guided reasoning distillation process. We employ Qwen2.5-32B-Instruct model as a teacher to generate structured vulnerability analyses for both vulnerable and non-vulnerable functions. 
Let 
\begin{equation}
\mathcal{D} = \{(x_i, y_i)\}_{i=1}^{N}
\end{equation}
denote a labeled dataset of $\
x_i$ with binary vulnerability labels $\
y_i \in \{\textsc{0}, \textsc{1}\}
$.
Following prior work on reasoning distillation for vulnerability detection \cite{weyssow2025r2vul, wen2025boosting}, we employ a teacher LLM to generate contrastive reasoning pairs for each function. For every sample, we construct:
\begin{itemize}
    \item a valid reasoning $\
r_i^{+}$, conditioned on the true label $\
y_i$
    \item a flawed reasoning $\
r_i^{-}$, obtained by swapping the label $\
y_i$
\end{itemize}
This yields a preference dataset: 
\begin{equation}
\mathcal{D}' = \{(x_i, y_i, r_i^{+}, r_i^{-})\}_{i=1}^{N}   
\end{equation}
The teacher is prompted to extract entities from the function (e.g., file paths, APIs, parameters), associate each entity with one or more vulnerability abstract classes, and map the reasoning to the CWE, supported by a concise vulnerability summary for vulnerable examples and summary highlighting the absence of known vulnerabilities for non-vulnerable examples. Each generated sample explicitly encodes an \textit{entity–class–CWE} alignment, grounding the explanation in structured security knowledge rather than free-form intuition. This structured alignment reduces ambiguity by distinguishing which specific entity–vulnerability relationships are actually implicated, rather than allowing the model to rely on diffuse or pattern-based intuition.

\subsection{Knowledge-Graph-guided Preference Optimization}
We fine-tune a student LLM using ORPO.
For each sample $\
(x_i;\, r_i^{+}, r_i^{-})$, we construct a preference pair: $\
(x_i;\, r_i^{+} \succ r_i^{-})$ where $\
r_i^{+}$	
  denotes the valid well-aligned with KG reasoning and $\
r_i^{-}$
  its flawed counterpart.
  ORPO combines a supervised fine-tuning (SFT) objective on $\
r_i^{+}$ , which teaches the model to generate structured and coherent reasoning; and a contrastive odds-ratio loss, which increases the log-odds of selecting $\
r_i^{+}$
  over $\
r_i^{-}$:

\begin{equation}
\mathcal{L}_{\text{ORPO}} =
\mathcal{L}_{\text{SFT}}(x_i, r_i^{+}; \theta)
+
\lambda \, \mathcal{L}_{\text{OR}}(x_i, r_i^{+}, r_i^{-}; \theta)
\end{equation}
where 
\begin{equation}
\mathcal{L}_{\text{SFT}}(x_i, r_i^{+}; \theta)
=
- \sum_{t} \log p_{\theta}(r_{i,t}^{+} \mid x_i, r_{i,<t}^{+})
\end{equation}

\begin{equation} \mathcal{L}_{\text{OR}} = -\frac{1}{n} \sum_{i=1}^{n} \log \sigma \!\left( \log \frac{\mathrm{odds}_{\theta}(r_i^{+} \mid x_i)} {\mathrm{odds}_{\theta}(r_i^{-} \mid x_i)} \right)\end{equation}

ORPO integrates preference learning directly into the supervised objective, yielding a simpler and more stable training procedure. In our setting, the odds-ratio term encourages the model to favor reasoning with entity–calss–CWE structure that consistent with KG, while suppressing superficially plausible but structurally incorrect explanations. This design encourages the model to internalize security concepts encoded in the KG, reinforcing causal reasoning over pattern memorization.
As a result, the model learns to generate explanations that are both accurate and semantically consistent with standardized vulnerability knowledge. Therefore, ORPO effectively addresses challenges by reducing semantic ambiguity and improving the reasoning performance of a model, limiting unconstrained reasoning that often leads to hallucinated or spurious security claims.

\subsection{Knowledge Graph Augmentation}
The KG used for distillation contains two primary concept types: abstraction classes and CWE nodes. We further enrich the KG iteratively by mining function-related entities from teacher-generated rationales, including function and variable names, API calls, libraries, and vulnerability indicators frequently associated with specific CWE categories. Newly mined entities are linked to existing abstraction classes and CWEs based on repeated co-occurrence patterns and semantic similarity. Over time, this augmentation integrates real-world code patterns into the KG, strengthening connections between abstract vulnerability definitions and concrete implementation artifacts.


\section{Experimental Setup}
\subsection{Research Questions}
This work aims to systematically evaluate the effectiveness of KG–guided LLMs for SVD. In particular, we focus on both coarse-grained vulnerability identification and fine-grained CWE classification, as well as the role of structured security knowledge in improving model reasoning and robustness. To this end, we formulate the following research questions:
\begin{itemize}
\item \textbf{RQ1: Overall Vulnerability Detection Performance.}
How does \tool perform in overall vulnerability detection compared to DL baselines and existing LLM-based classifiers under binary SVD settings?
\item \textbf{RQ2: CWE-level Reasoning.}
How effective is \tool in generating CWE-consistent reasoning, and how does extracting CWE identifiers from these reasonings translate into accurate multi-class vulnerability classification across large and long-tailed CWE spaces?
\item \textbf{RQ3: KG-Guidance Contribution.}
To what extent does KG grounding — through CWE abstraction and structured entity relations — improve both binary detection (RQ1) and CWE-level reasoning (RQ2), in terms of accuracy and reliability?
\end{itemize}

\subsection{Datasets}

\subsubsection{Dataset Overview}
We evaluate \tool on three recent large-scale software vulnerability datasets: \textbf{DiverseVul}~\cite{chen2023diversevul}, \textbf{PrimeVul}~\cite{ding2024vulnerability}, and \textbf{R2Vul}~\cite{weyssow2025r2vul} as summarized in Table \ref{tab:dataset_stats}. These datasets are selected for their complementary strengths in vulnerability coverage, data quality, and real-world grounding. All samples are derived from real GitHub commits, ensuring that the evaluated vulnerabilities reflect practical development scenarios rather than synthetic patterns. Moreover, all three datasets were released between 2023 and 2024, making them representative of contemporary software vulnerabilities.

The datasets differ substantially in their CWE coverage, collectively spanning a broad and challenging multi-class classification space. DiverseVul covers approximately 150 distinct CWEs, PrimeVul includes around 140 CWEs, and R2Vul extends this coverage to 267 CWEs, introducing a long-tail distribution that is particularly challenging for fine-grained vulnerability classification. This diversity allows us to assess both binary vulnerability detection and CWE-level reasoning under realistic and heterogeneous conditions.

\begin{table}[t]
\centering
\caption{Dataset Statistics.}
\label{tab:dataset_stats}
\begin{tabular}{lccc}
\toprule
Dataset & Function & CWEs & Splits \\
\midrule
DiverseVul~\cite{chen2023diversevul} & 349,437 & 150 & 14400 / 1800  / 1800 \\
R2Vul~\cite{weyssow2025r2vul}     & 235,768 & 267 & 13829 / 1734 / 1734 \\
PrimeVul~\cite{ding2024vulnerability}  & 18000 & 140 & 14400 / 1800 / 1800 \\
\bottomrule
\end{tabular}
\end{table}

\subsubsection{Preprocessing and Dataset Balancing}
To ensure comparability across datasets, we align all benchmarks to a common scale of approximately 18000 samples, matching the size of R2Vul corpus. Since DiverseVul and PrimeVul exhibit severe class imbalance (with vulnerable-to-non-vulnerable ratios of approximately 1:17 and 1:33, respectively), we apply downsampling to the majority class while preserving diversity across CWE categories. Samples without explicit CWE identifiers are excluded to enable consistent CWE-level reasoning and evaluation.

For preference-based training, all three datasets are further enriched with KG-guided valid and flawed reasoning pairs generated by the teacher Qwen2.5-32B-Instruct model as described in Section \ref{distill}. Each dataset is split into training, validation, and test sets using an 8:1:1 ratio. These splits are used consistently across SFT, ORPO training, and evaluation to ensure fair comparison.

All three datasets are employed across our RQs with the same preprocessing and balancing procedures applied to maintain consistency. However, for subset-aligned evaluations in Section \ref{sec:subsetres}, we follow the settings of prior binary SVD baselines. Specifically, PrimeVul is the only dataset shared across all methods, as both LLMxCPG~\cite{lekssays2025llmxcpg} and ReVD~\cite{wen2025boosting} derive task-specific training data from PrimeVul and do not provide compatible splits for other datasets considered in this work. Reconstructing their dataset distillation procedures across additional benchmarks would require substantial re-engineering and is beyond the scope of this study. Nevertheless, evaluating all methods on PrimeVul ensures a controlled and meaningful comparison under identical test conditions.

\subsection{Baselines}
\subsubsection{Traditional Deep Learning Baselines}
We compare against three representative SVD models that do not rely on LLMs. A BiLSTM-based sequence model is included as a token-level baseline following the formulation of prior work on vulnerability detection \cite{farasat2024machine}. SySeVR is used as a semantic-based baseline that leverages code slicing and vulnerability-specific representations to capture security-relevant semantics \cite{li2021sysevr}. We additionally include Devign, a graph-based model that integrates control-flow and data-flow information through graph neural networks \cite{zhou2019devign}. Together, these baselines represent sequence-, semantic-, and graph-centric paradigms in traditional vulnerability detection. They are evaluated in both binary (RQ1) and multi-class (RQ2) settings, but do not generate natural-language reasoning text.

\subsubsection{LLM-Based Models}
For LLM-based evaluation, we consider three instruction-tuned code-oriented language models from different model families: Qwen2.5-7B-Instruct, LLaMA-3.1-8B-Instruct, and deepseek-coder-7b-instruct-v1.5 models of comparable scale. Each model is evaluated under four adaptation strategies: zero-shot prompting, in-context learning (ICL), chain-of-thought prompting that encourages step-by-step reasoning (COT), and SFT using Quantized Low-Rank Adaptation (QLoRA). For consistency and fairness, prompting templates and fine-tuning configurations follow the settings of the prior work \cite{bi2023benchmarking}. These serve as student baselines for our \tool in RQ1 and RQ2. Since these models generate textual explanations, CWE predictions in the multi-class setting are derived by extracting CWE identifiers from their reasoning text.

\subsubsection{State-of-the-art SVD Baselines}
In addition, as SVD baselines we compare out approach against three recent LLM-based SVD approaches: R2Vul~\cite{weyssow2025r2vul}, ReVD~\cite{wen2025boosting}, and LLMxCPG~\cite{lekssays2025llmxcpg}. These baselines represent state-of-the-art approaches that enhance LLM vulnerability detection through structured reasoning, curriculum-based preference optimization, or code property graph guidance. 
For RQ1, to enable a fair comparison we conduct subset-aligned evaluations against LLMxCPG and ReVD under their respective experimental assumptions. Both rely on custom preprocessing and training pipelines, which differ substantially from standard multi-dataset or multi-CWE evaluation settings; therefore, they are not included in RQ2 multi-class comparisons. R2Vul, on the contrary, is evaluated in RQ2 alongside our approach and traditional DL baselines because of its reasoning nature.

\subsection{Evaluation Metrics}
We evaluate model performance under both binary and multi-class vulnerability detection settings, corresponding to RQ1 and RQ2, respectively. For RQ1, we report Precision (P), Recall (R), and F1-score, which are standard metrics in SVD and capture the trade-off between false positives and false negatives. Precision measures the proportion of predicted vulnerabilities that are correct, recall measures the proportion of true vulnerabilities that are successfully identified, and F1-score represents their harmonic mean. For RQ2 results, we report Micro-F1 and Macro-F1 scores. Micro-F1 aggregates contributions from all classes and is dominated by frequent CWEs, reflecting overall classification accuracy. Macro-F1 computes the unweighted average F1-score across all CWEs, treating rare and frequent classes equally. In addition, we report Micro and Macro Precision/Recall: Micro-Precision and Micro-Recall emphasize performance on frequent CWE categories by weighting results according to class frequency, while Macro-Precision and Macro-Recall average scores across all classes, providing a balanced view of performance on both common and rare CWEs. This metric is particularly important for assessing performance on long-tail CWEs, which are prevalent in real-world vulnerability datasets and central to our evaluation objectives. For LLM-based baselines, CWE predictions in the multi-class setting are derived by extracting CWE identifiers directly from the generated reasoning text. Thus, reported Micro-F1 and Macro-F1 scores reflect reasoning accuracy, i.e., the model’s ability to correctly identify the ground-truth CWE within its explanation. In contrast, DL baselines do not generate reasoning and are evaluated solely on their classifier outputs. Finally, in RQ3 we assess the impact of KG by comparing performance with and without KG integration under the same evaluation metrics.

\subsection{Implementation Details}
All LLM-based experiments employ fine-tuning using QLoRA within the PEFT framework. Models are quantized to 4-bit precision, with a LoRA rank of 16 and scaling factor of 32. Optimization is performed using AdamW with a learning rate of $2 \times 10^{-5}$ and a batch size of 64. Only adapter parameters are updated during training, while the base model weights remain frozen.
We use a maximum input length of 4096 tokens and apply greedy decoding during inference with sampling disabled. All experiments are conducted on NVIDIA A100 GPUs with 80GB memory and RTX-class GPUs with 40GB memory, using mixed-precision training with FP16 enabled. Model training and evaluation are implemented using HuggingFace’s Trainer API.

\section{Results and Discussion}
\subsection{RQ1: Overall Vulnerability Detection Performance}

\begin{table*}[!htp]
\centering
\caption{Binary performance comparison with Baselines. \underline{Underlined} values indicate the best performance within each model family. \textbf{Bold} values indicate the best performance across all models for a given dataset and metric.}
\label{tab:rq1}
\scalebox{0.925}{
\begin{tabular}{llccccccccc}
\toprule
\multicolumn{2}{c}{\multirow{2}{*}{Baselines}}
& \multicolumn{3}{c}{DiverseVul} 
& \multicolumn{3}{c}{R2Vul} 
& \multicolumn{3}{c}{PrimeVul} \\
\cmidrule(lr){3-5} \cmidrule(lr){6-8} \cmidrule(lr){9-11}
& & Precision & Recall & F1 & Precision & Recall & F1 & Precision & Recall & F1\\
\midrule

\multirow{3}{*}{}
& BiLSTM & \underline{0.50} & \underline{0.70} & \underline{0.59} & \underline{0.55} & \underline{0.72} & \underline{0.62} & \underline{0.45} & \underline{0.68} & \underline{0.55} \\
& SySeVR & 0.15 & 0.25 & 0.19 & 0.08 & 0.15 & 0.11 & 0.12 & 0.20 & 0.15 \\
& Devign & 0.20 & 0.30 & 0.24 & 0.22 & 0.32 & 0.26 & 0.14 & 0.22 & 0.17 \\

\midrule

\multirow{5}{*}{LLaMA}
& Zero-Shot & 0.22 & 0.55 & 0.31 & 0.28 & 0.64 & 0.39 & 0.21 & 0.44 & 0.28 \\
& ICL & 0.30 & 0.64 & 0.41 & 0.39 & 0.61 & 0.48 & 0.33 & 0.50 & 0.41 \\
& COT & 0.36 & \underline{0.76} & 0.49 & 0.41 & 0.73 & 0.53 & 0.39 & 0.51 & 0.44 \\
& SFT & 0.45 & 0.63 & 0.53 & 0.52 & 0.68 & 0.59 & 0.40 & 0.58 & 0.48 \\
& \textbf{\tool} & \underline{0.66} & 0.68 & \underline{0.67} & \underline{0.67} & \underline{0.69} & \underline{0.68} & \underline{0.57} & \underline{0.59} & \underline{0.58} \\

\midrule

\multirow{5}{*}{DeepSeek}
& Zero-Shot & 0.15 & 0.66 & 0.24 & 0.18 & 0.70 & 0.29 & 0.20 & 0.69 & 0.31 \\
& ICL & 0.21 & 0.75 & 0.39 & 0.26 & \underline{0.75} & 0.39 & 0.26 & 0.74 & 0.38 \\
& COT & 0.23 & \underline{0.79} & 0.36 & 0.32 & 0.74 & 0.48 & 0.28 & \underline{0.80} & 0.41 \\
& SFT & 0.42 & 0.63 & 0.51 & 0.44 & 0.63 & 0.52 & 0.41 & 0.64 & 0.51 \\
& \textbf{\tool} & \underline{0.64} & 0.66 & \underline{0.65} & \underline{\textbf{0.69}} & 0.71 & \underline{0.70} & \underline{0.59} & 0.61 & \underline{0.60} \\

\midrule

\multirow{5}{*}{Qwen}
& Zero-Shot & 0.25 & 0.50 & 0.33 & 0.21 & 0.63 & 0.32 & 0.30 & 0.72 & 0.42 \\
& ICL & 0.37 & 0.67 & 0.48 & 0.41 & 0.72 & 0.52 & 0.35 & \underline{\textbf{0.77}} & 0.48 \\
& COT & 0.38 & 0.60 & 0.47 & 0.46 & 0.75 & 0.57 & 0.36 & 0.75 & 0.48 \\
& SFT & 0.50 & 0.73 & 0.60 & 0.58 & 0.79 & 0.67 & 0.40 & 0.62 & 0.49 \\
& R2Vul & 0.56 & 0.72 & 0.63 & 0.56 & \underline{\textbf{0.90}} & 0.69 & 0.50 & 0.68 & 0.58 \\
& \textbf{\tool} & \underline{\textbf{0.78}} & \underline{\textbf{0.80}} & \underline{\textbf{0.79}} & \underline{0.68} & 0.88 & \underline{\textbf{0.77}} & \underline{\textbf{0.69}} & 0.67 & \underline{\textbf{0.68}} \\

\bottomrule
\end{tabular}
}
\end{table*}

Table~\ref{tab:rq1} summarizes binary vulnerability detection performance across all evaluated baselines and shows that LLM-based methods consistently outperform deep learning (DL) baselines on all three datasets. Overall, DL models achieve substantially lower F1-scores (e.g., 0.11--0.26 on R2Vul and 0.15--0.24 on PrimeVul), indicating limited generalization beyond surface patterns and weaker robustness across benchmarks. In contrast, general-purpose LLMs without task-specific training already provide a clear improvement, with zero-shot and in-context variants reaching F1-scores in the 0.33--0.48 range. SFT further increases performance across datasets, achieving F1-scores up to 0.67 on DiverseVul and R2Vul.

Across all settings, \tool achieves the strongest results and consistently outperforms both DL and LLM-based baselines. When integrated with Qwen, \tool attains the highest F1-scores of 0.79, 0.77, and 0.68 on DiverseVul, R2Vul, and PrimeVul, respectively, surpassing SFT as well as the reasoning-based baseline R2Vul. We observe similar gains for LLaMA and DeepSeek, suggesting that the benefits of KG guidance are not model-specific but transfer across different student backbones. Taken together, these results indicate that incorporating structured security knowledge improves the accuracy and stability of binary vulnerability detection across diverse real-world benchmarks.

\paragraph{Subset-Aligned Comparison with LLMxCPG and ReVD}
\label{sec:subsetres}
To ensure fair comparison, we evaluated \tool against LLMxCPG and ReVD under their respective experimental settings using the PrimeVul dataset.
For LLMxCPG, we restricted evaluation to the CWEs supported in its original setup and adopted its reported splits. As shown in Table \ref{tab:rq1llmxcpg}, our method achieved an F1-score of 71.4\% compared to 67.2\% for LLMxCPG, with improvements in recall for 12.5\% while maintaining balanced precision.
For ReVD, we used its original PrimeVul splits. Table \ref{tab:rq1revd} shows that \tool achieves an F1-score of 64\% compared to 53\% for ReVD, improving precision and recall for 7\% and 2\%, respectively.
Across both comparisons, \tool consistently outperformed state-of-the-art binary SVD methods, demonstrating improved vulnerability detection under constrained evaluation settings.

\begin{table}[!htp]
\centering
\begin{minipage}[t]{0.48\linewidth}
\centering
\caption{Binary performance under restricted CWE subsets with LLMxCPG.}
\label{tab:rq1llmxcpg}
\begin{tabular}{lccc}
\toprule
Model & Precision & Recall & F1 \\
\midrule
LLMxCPG & \textbf{86.3}\% & 55\% & 67.2\% \\
\textbf{\tool} & 83.3\% & \textbf{62.5}\% & \textbf{71.4}\% \\
\bottomrule
\end{tabular}
\end{minipage}\hfill
\begin{minipage}[t]{0.48\linewidth}
\centering
\caption{Binary performance comparison on PrimeVul using ReVD's original splits.}
\label{tab:rq1revd}
\begin{tabular}{lccc}
\toprule
Model & Precision & Recall & F1 \\
\midrule
ReVD & 58\% & 64\% & 53\% \\
\textbf{\tool} & \textbf{65}\% & \textbf{66}\% & \textbf{64}\% \\
\bottomrule
\end{tabular}
\end{minipage}
\end{table}

\begin{tcolorbox}[colback=gray!20, colframe=black, boxrule=0.2mm, rounded corners]
Answer to RQ1: \tool consistently outperformed DL baselines and existing LLM methods under binary SVD settings, achieving the highest F1-scores across diverse datasets and model families.
\end{tcolorbox}
\subsection{RQ2: CWE-level Reasoning}
\subsubsection{LLM Reasoning-based CWE Classification}
Table \ref{tab:rq2} reports Macro-F1 and Micro-F1 scores for LLM-based models. Since these models produce reasoning text, we extract CWE identifiers from their reasoning and compare them with ground-truth CWE labels in dataset. Thus, the reported classification metrics directly reflect the reasoning accuracy.
\tool consistently achieves the highest scores across all datasets. In DiverseVul, it reaches 34.4\% Macro-F1 and 23.4\% Micro-F1, outperforming other LLM baselines by a clear margin. Similar improvements are observed with R2Vul dataset with 29.9\% Macro-F1 and 25.2\% Micro-F1 and 35.6\% Macro-F1 and 22.3\% Micro-F1 over PrimeVul. These results demonstrate that \tool not only predicts CWE classes more reliably but also generates rationales that correctly identify CWE with higher accuracy.
\commentout{
\begin{table*}[!htp]
\centering
\caption{Multi-Class CWE Classification.}
\label{tab:rq2}
\begin{tabular}{lcccccc}
\toprule
\multirow{2}{*}{Baseline} 
& \multicolumn{2}{c}{DiverseVul} 
& \multicolumn{2}{c}{R2Vul} 
& \multicolumn{2}{c}{PrimeVul} \\
\cmidrule(lr){2-3} \cmidrule(lr){4-5} \cmidrule(lr){6-7}
& Macro-F1 & Micro-F1 & Macro-F1  & Micro-F1  & Macro-F1  & Micro-F1  \\
\midrule
SFT (LLaMA)     & 0.021 & 0.043 & 0.032 & 0.044 & 0.023 & 0.050 \\
\textbf{\tool (LLaMA)}     & \textbf{0.313} & \textbf{0.192} & \textbf{0.208} & \textbf{0.187} & \textbf{0.337} & \textbf{0.207} \\
\midrule
SFT (DeepSeek)  & 0.033 & 0.018 & 0.020 & 0.064 & 0.026 & 0.057 \\
\textbf{\tool (DeepSeek)}  & \textbf{0.331} & \textbf{0.188} & \textbf{0.230} & \textbf{0.219} & \textbf{0.359} & \textbf{0.213} \\
\midrule
SFT (Qwen)      & 0.019 & 0.057 & 0.028 & 0.047 & 0.031 & 0.049 \\
R2Vul (Qwen)    & 0.030 & 0.063 & 0.028 & 0.067 & 0.036 & 0.059 \\
\textbf{\tool (Qwen)}    & \textbf{0.344} & \textbf{0.234} & \textbf{0.299} & \textbf{0.252} & \textbf{0.356} & \textbf{0.223} \\
\bottomrule
\end{tabular}
\end{table*}
}
\begin{table*}[!htp]
\centering
\caption{Multi-Class CWE Classification.}
\label{tab:rq2}
\begin{tabular}{llcccccc}
\toprule
\multicolumn{2}{c}{\multirow{2}{*}{Baseline}} &
\multicolumn{2}{c}{DiverseVul} &
\multicolumn{2}{c}{R2Vul} &
\multicolumn{2}{c}{PrimeVul} \\
\cmidrule(lr){3-4} \cmidrule(lr){5-6} \cmidrule(lr){7-8}
 &  & Macro-F1 & Micro-F1 & Macro-F1 & Micro-F1 & Macro-F1 & Micro-F1 \\
\midrule
\multirow{2}{*}{LLaMA} 
    & SFT & 0.021 & 0.043 & 0.032 & 0.044 & 0.023 & 0.050 \\
    & \textbf{\tool} & \textbf{0.313} & \textbf{0.192} & \textbf{0.208} & \textbf{0.187} & \textbf{0.337} & \textbf{0.207} \\
\midrule
\multirow{2}{*}{DeepSeek} 
    & SFT & 0.033 & 0.018 & 0.020 & 0.064 & 0.026 & 0.057 \\
    & \textbf{\tool} & \textbf{0.331} & \textbf{0.188} & \textbf{0.230} & \textbf{0.219} & \textbf{0.359} & \textbf{0.213} \\
\midrule
\multirow{3}{*}{Qwen} 
    & SFT & 0.019 & 0.057 & 0.028 & 0.047 & 0.031 & 0.049 \\
    & R2Vul & 0.030 & 0.063 & 0.028 & 0.067 & 0.036 & 0.059 \\
    & \textbf{\tool} & \textbf{0.344} & \textbf{0.234} & \textbf{0.299} & \textbf{0.252} & \textbf{0.356} & \textbf{0.223} \\
\bottomrule
\end{tabular}
\end{table*}

\subsubsection{DL Multi-Class Performance}
\label{dlmulti}
This section summarizes the performance of BiLSTM, SySeVR, Devign approaches trained for CWE-level prediction. Unlike LLMs, these models do not generate reasoning and are evaluated solely on the classification output. Their performance is extremely limited, with Macro-F1 scores averaging below 0.0005 and Micro-F1 scores below 0.001 across all datasets. BiLSTM achieves the best results and is the only baseline reported in Table \ref{tab:bilstm_cwe}, as SySeVR and Devign perform near-zero, preventing meaningful comparison.
Although DL baselines provide a reference point for direct CWE classification, their inability to capture semantic and structural diversity highlights the advantage of reasoning-driven LLM approaches.
\begin{tcolorbox}[colback=gray!20, colframe=black, boxrule=0.2mm, rounded corners]
Answer to RQ2: \tool outperforms all LLM baselines in CWE-consistent reasoning and DL baselines with near-zero performance achieving the highest Macro-F1 and Micro-F1 scores across all datasets.
\end{tcolorbox}
\subsection{RQ3: KG-Guidance Contribution}

We further compared best performing \tool on the R2Vul dataset. For the baseline without KG, we adopted the original prompt design from the \cite{weyssow2025r2vul} to ensure fairness. As shown in Table \ref{tab:rq3}, KG implementation shows stronger results in both binary detection and CWE-level reasoning.
In binary detection, \tool improves precision, recall, and F1-score, reaching 0.67, 0.88, and 0.77 with KG integration respectively, compared to 0.52, 0.84, and 0.64 for \tool without KG. In reasoning evaluation, the gains are more pronounced: Macro-F1 rises to 0.30 from 0.05, while Micro-F1 increases to 0.25 from 0.07 when model uses KG.
These results confirm that integrating KG reasoning substantially enhances both vulnerability detection and CWE identification. The improvements highlight the role of structured knowledge in guiding reasoning, enabling more accurate classification and reducing errors.

\begin{table*}[!htp]
\centering
\caption{Impact of KG.}
\label{tab:rq3}
\begin{tabular}{lccccc}
\toprule
& Precision & Recall & F1-score & Macro-F1 & Micro-F1 \\
\midrule
W/o KG             & 0.52 & 0.84 & 0.64 & 0.05 & 0.07 \\
\textbf{With KG}   & \textbf{0.67} & \textbf{0.88} & \textbf{0.77} & \textbf{0.30} & \textbf{0.25} \\
\bottomrule
\end{tabular}
\end{table*}

\begin{tcolorbox}[colback=gray!20, colframe=black, boxrule=0.2mm, rounded corners]
Answer to RQ3: Our findings confirm that integrating KG grounding through CWE abstraction and structured relations significantly improves binary detection and multi‑class CWE reasoning, yielding higher F1, Macro‑F1, and Micro‑F1 scores and enhancing overall accuracy and reliability in vulnerability detection.
\end{tcolorbox}

\subsection{Discussion}

\subsubsection{DL baselines}
DL Baselines in Section \ref{dlmulti} exhibited poor performance in RQ2, particularly in multi-class CWE classification. In this section we discuss additional experiment that show improvements on BiLSTM baseline. Our training corpus contains over 200 CWE types in R2Vul corpus, which introduced a severe imbalance across classes. We experimented with several standard strategies to address this issue, including downsampling, hybrid sampling, and class-weight adjustments. None of these approaches yielded meaningful improvements, and the model consistently collapsed to predicting a single CWE with negligible precision.
To further investigate, we restricted the classification task to a smaller subset of the data. Specifically, we selected the first 20 CWE types from the MITRE Top-25 list and trained the BiLSTM on this reduced set. Under this setting, the model achieved measurable improvements, reaching a Macro Precision of 0.0502, Macro Recall of 0.1154, and Macro F1-score of 0.0641, compared to near-zero values in the full R2Vul setting as showed in Table \ref{tab:bilstm_cwe}. While these scores remain modest, they demonstrate that limiting the classification space to frequent and well-represented CWEs can partially mitigate the shortcomings of DL baselines.

\begin{table*}[t]
\centering
\caption{BiLSTM Performance under Different CWE Subset Training Settings.}
\label{tab:bilstm_cwe}
\begin{tabular}{lccc}
\toprule
Training Setting & Macro Precision & Macro Recall & Macro F1-score \\
\midrule
BiLSTM (Full CWE subset)   & 0.0136 & 0.0002 & 0.0004 \\
BiLSTM (Top-25 CWE subset) & \textbf{0.0502} & \textbf{0.1154} & \textbf{0.0641} \\
\bottomrule
\end{tabular}
\end{table*}

\subsubsection{Impact of KG retrieval}
The ablation study in RQ3 highlights that KG guidance substantially improves both binary detection accuracy and CWE-level reasoning. This distinction reflects the nature of the task: binary detection relies primarily on surface-level vulnerability cues, whereas CWE attribution requires structured semantic alignment. By retrieving knowledge from the KG, we enable reasoning that is more precise and less prone to confusion between related categories. These findings confirm that KG integration is most critical for fine-grained classification rather than coarse detection.
\subsubsection{Impact of Knowledge Guidance}
We examine the effect of different information sources on model reasoning. In the original R2Vul setup, reasoning traces often contained CVE identifiers. Our analysis showed that these identifiers were consistently hallucinated and did not match ground-truth labels. When included, they negatively affected performance of \tool. For this reason, we masked CVE-IDs during training to prevent the model from relying on hallucinating signals.
We also evaluated the role of entities and abstract class information derived from the KG. These features did not improve binary detection when used in isolation. However, they contributed to better CWE-level reasoning when combined with KG guidance. This indicates that such information is useful only in contexts where structured relational knowledge is available to support reasoning.

\section{Threats to Validity}
\paragraph{CWE classification}
Our KG organizes CWEs into 13 abstract classes inspired by MITRE documentation and prior work \cite{li2025safegenbench}. However, this taxonomy is not fixed and alternative groupings may be more precise or less granular. The assignment of CWEs to these classes is based on keyword and behavioral cues, which introduces the possibility of misclassification. As a result, the reported improvements in reasoning may depend on a chosen abstraction scheme.
\paragraph{Data Distillation}
The development of our baselines relied on a large corpus of training data generated with the assistance of a teacher LLM. While this process aimed to ensure coverage and consistency, we cannot guarantee the absolute accuracy of every reasoning pair in the dataset. Errors or inconsistencies in the generated rationales may propagate into the trained models and influence evaluation outcomes.
\paragraph{Retrieved KG Context}
At inference time, KG retrieval occasionally returns extensive contextual information, including multiple CWE candidates associated with extracted entities. In such cases, the model may mention several CWEs and conclude that the function could relate to any of them. To mitigate this, we incorporated confidence levels based on entity frequency, but this approach remains limited. Future work will explore more refined retrieval strategies to reduce noise and improve precision.
\paragraph{Computational constraints}
The number of hyperparameter trials for QLoRA configuration and the LLM generation method was relatively limited. Additionally, due to GPU constraints, we did not attempt to fine-tune larger models, such as the 34B version. Although these limitations may have resulted in alternative findings, they do not fundamentally undermine our primary objective: to explore the potential of conducting vulnerability detection and explanation tasks using LLMs.
\section{Related Work}
\subsection{Vulnerability Detection}
DL models have been widely applied to software vulnerability detection \cite{zhou2019devign}. These approaches achieve reasonable performance in binary classification tasks but operate as black-box predictors, offering little insight into the underlying causes of vulnerabilities.  Improving those architectures with representative learning improved overall performance, however those approaches still are limited to labels, without any reasoning or explanation that could support interpretation or remediation \cite{wen2024livable, li2018vuldeepecker, hin2022linevd, wartschinski2022vudenc, farasat2024machine, rahman2024towards, cao2024snopy}.

With the emergence of large language models, researchers began exploring prompting and fine-tuning strategies to apply LLMs to vulnerability detection \cite{sheng2024lprotector, zhang2024prompt, ding2024vulnerability, yang2025dlap, gonccalves2024scope, nong2022open, tamberg2025harnessing}. These models leverage natural language understanding and generation capabilities to produce diverse outputs, including textual rationales or fix suggestions, in addition to binary detection. More recent studies have also investigated preference optimization methods to align reasoning quality \cite{wen2025boosting}. While these approaches improve surface-level coherence, they remain constrained to binary classification and do not reliably connect explanations to specific vulnerability categories.

There are also works that attempt to extend SVD models with external knowledge, such as textual summaries, graphs, or domain-specific attributes \cite{sheng2024lprotector, nguyen2025safe, lekssays2025llmxcpg, du2024vul, zheng2025learning, sun2024llm4vuln,wang2024llmdfa}. These methods provide additional context and yield improvements in certain cases, but continue to evaluate performance primarily in terms of binary detection \cite{deng2024improving}. Some studies have experimented with vulnerability type classification, but typically cover only a limited set of CWE categories and binary classification performance within specific vulnerability types \cite{wen2024livable, gao2023far,mirsky2023vulchecker}.

A smaller number of approaches have explored generating explanations alongside detection. However, these explanations are often vague, not explicitly tied to the vulnerability type or perform detection over limited classes \cite{ullah2024llms, yin2024multitask, mao2024towards, ni2023distinguishing, steenhoek2024err, zibaeirad2025reasoning, yang2025dlap, li2024iris, jararweh2025llavul}. R2Vul is notable in this regard, as it introduces reasoning aligned with CWE identifiers \cite{weyssow2025r2vul}. Yet, our analysis shows that the generated reasoning is frequently incorrect or misclassified, limiting its reliability. This highlights a broader challenge: while LLMs can generate text that resembles reasoning, the correctness and usefulness of that reasoning for vulnerability analysis remain uncertain \cite{yu2024insight,wang2024llmdfa}.

Against this background, our work investigates whether LLMs can truly reason about vulnerabilities in source code functions, and whether their reasoning ability can be effectively supported to improve CWE attribution.

\subsection{Knowledge Graphs in the Vulnerability Domain}

Vulnerability KGs are primarily used to support high-level security analysis and decision-making. Common applications include vulnerability retrieval and similarity search (e.g., a multi‑faceted search website \cite{sun2023multi} and an intelligent retrieval system for similar information‑system vulnerabilities \cite{dong2024intelligent}), entity alignment across heterogeneous databases (the Neighborhood Matching model that aligns CVE, China National Vulnerability Database\footnote{https://www.cnnvd.org.cn} (CNVD) and other sources \cite{yan2023neighborhood}), risk assessment and co‑exploitation analysis (the compact VulKG used for vulnerability‑co‑exploitation behavior discovery \cite{yin2024compact}), and attack‑path or link‑prediction modeling (text‑enhanced Graph Attention Networks combined with LLM reasoning to infer missing relations \cite{liu2025dynamic}). Empirical evaluations consistently show that KG-based approaches outperform non-graph-based baselines on these tasks, enabling more accurate retrieval, stronger relational inference, and better integration of heterogeneous security data.
A key strength of KGs lies in their ability to represent contextual and relational information explicitly by modeling relationships between vulnerabilities, weaknesses, products, and attacks \cite{wu2025graph, wang2025knowledge}. KGs capture dependencies and patterns that are difficult to express using flat feature representations alone \cite{liu2025dynamic, yin2024compact}.

Despite their demonstrated benefits, existing vulnerability KGs are rarely used in the context of SVD. Previous work mainly treats KGs as repositories for organizing, retrieving, or reasoning over known vulnerabilities, rather than as components that actively guide detection models \cite{wang2024research}.
Moreover, most studies rely on similarity-based retrieval or graph embeddings, while the role of explicit semantic relations—such as CWE hierarchies or security abstractions—remains largely unexplored. As a result, there is limited understanding of how structured vulnerability knowledge can be used to improve model reasoning, reduce inconsistent explanations, or enforce alignment with established vulnerability taxonomies.

\section{Conclusion}

This paper introduces \tool, a knowledge-graph–guided approach to software vulnerability reasoning and detection. By injecting structured security knowledge into dataset construction, training, and inference, \tool supports CWE-level reasoning over long-tailed vulnerability distributions. We also propose KG-informed dataset distillation and an ORPO-based preference optimization method with parameter-efficient fine-tuning to improve the factual accuracy and CWE consistency of LLM explanations.
Empirically, we find that LLMs outperform deep learning baselines for binary SVD, with supervised fine-tuning performing best. For CWE-level reasoning, \tool substantially improves semantic consistency and accuracy over SFT-only LLMs and the R2Vul baseline. Ablations further show KG guidance is critical, especially for multi-class settings, improving both detection and reasoning reliability.
We note validity threats from non-unique CWE abstractions, teacher-generated rationale noise, and imperfect KG retrieval, and our exploration is limited by compute. Future work will refine CWE taxonomies and retrieval, scale to larger models, and extend KG-guided reasoning to runtime analysis and broader security tasks.

\section{Data Availability}
Source code and data are available at \url{https://anonymous.4open.science/r/Vul-ReaD}.

\bibliographystyle{ACM-Reference-Format}
\bibliography{sample-base}

\appendix

\end{document}